# Metrics for Video Quality Assessment in Mobile Scenarios

Gaurav Pande
Department of MCA,
RK Goel Institute of Technology
Ghaziabad, India
reachgauravpande@gmail.com

**Abstract.** With exponential increase in the volumes of video traffic in cellular networks, there is an increasing need for optimizing the quality of video delivery. 4G networks (Long Term Evolution – Advanced or LTE-A) are being introduced in many countries worldwide, which allow a downlink speed of upto 1 Gbps and uplink of 100 Mbps over a single base station. This makes a strong push towards video broadcasting over LTE networks, characterizing its performance and developing metrics which can be deployed to provide user feedback of video quality and feed-back them to network operators to fine-tune the network. In this paper, we characterize the performance of video transmission over LTE-A physical layer using popular video quality metrics such as SSIM, Blocking, Blurring, NIQE and BRISQUE. We conduct experiments to find a suitable no-reference metrics for mobile scenario and find that Blocking Metrics is most promising in case of channel or modulation variations but it doesn't perform well to quantize variations in compression ratios. The metrics BRISQUE is very efficient in quantizing this distortion and performs well in case of network variations also.

## 1    Introduction

Mobile video is quickly becoming a mass consumer phenomenon, much as digital photos were earlier in the smartphone adoption cycle. There is a tectonic shift to on-demand and on-the-go video services. Video streaming services such as Netflix and Youtube are generating large volumes of traffic and revenue (3 billion $ for Netflix in 2011).

Moreover, the cellular networks are migrating towards a new technology 4G LTE which is 10x faster than 3G. Boosted bandwidth allows for quicker uploads and better streaming quality. LTE-A targets to achieve peak uplink and downlink data rates of 500 Mbps and 1 Gbps, respectively, for low-speed UEs and around 100 Mbps for those with higher mobilities. It accommodates the next generation of telecommunication services such as realtime high-definition video streaming, mobile HDTV and high quality video conferencing. In LTE-A systems, the bandwidths in both uplink and downlink can go upto 100 MHz, which is achieved by Carrier Aggregation (CA) or aggregation of individual Component Carriers

There have been volumes of research in different aspects of multimedia systems, to enhance the performance of mobile video. Hardware architectures have been proposed for mobile video compression and encryption [9-11]. Efficient video transmission in cellular low bandwidth scenarios has also been discussed [12-16]. With 4G LTE becoming popular, we attempt to understand the features of LTE which help in high quality video transmission.

The major features that distinguish LTE from 3G technologies at the air-interface are Orthogonal Frequency Multiple Access (OFDMA), advanced MIMO technology, and Hybrid Automatic Repeat Request (HARQ). In addition LTE uses flat-IP architecture for the core network. LTE uses OFDMA in the downlink (DL) for efficient multiple-access and for countering multipath frequency selective fading. OFDMA divides the available channel into number of sub-carriers and is naturally suitable for scalable bandwidth allocation by varying the FFT (Fast Fourier Transform) size. LTE uses SC-FDMA (Single Carrier FDMA) [6] in the uplink (UL) to obtain a low peak-to-average-power ratio (PAPR). In this paper, we concentrate only on the DL. LTE supports a full range of multiple antenna transmission techniques including transmit diversity (TD) [2], spatial multiplexing (SM) [3], and closed-loop eigen-beamforming [4] that are suited for different objectives. Transmit diversity is used for obtaining reliable transmissions and is achieved by using Space Frequency Block Codes [5] in LTE. SM is used for obtaining enhanced throughput and is achieved by using layered space time codes [3]. LTE also provisions for Multimedia Broadcast and Multicast through dedicated channels (called eMBMS services) and is becoming an attractive choice for next generation video broadcast.

With all these new features, the growth of mobile videos is on increase in wireless and mobile networks. However, there are not many (any) tools to measure the quality of received image/ video on-the-fly. We need metrics which can robustly indicate the quality of video received by the end user, don't drain the battery of smartphone and can give quality scores in real-time, without the need of reference video at the source. To this end, in this paper, we select some most common metrics used for image/ video quality assessment in the research literature and compare the performance of these metrics with each other. We choose three scenarios to validate the performance of these metrics:
   a) Changing Network conditions, where the channel SNR is varying from 10 to 20 dbs, keeping the modulation fixed.
   b) Changing modulation rate, where channel is kept fixed and modulation chosen by the scheme is changed.
   c) Changing source codec parameters.

We choose two full-reference metrics (which require original image/ video frame) PSNR and SSIM and four popular no-reference metrics namely Blocking, Blurring, NIQE and BRISQUE in our experiments. PSNR is the most popular and used as a reference in our experiments. Although, PSNR is sometimes found to be not robust in case of multiple distortion datasets, in our case, we apply a single distortion only, hence the choice of PSNR as the reference is justified.

The paper is organized as follows: Section 2 gives an overview of the LTE simulator we used in our experiments. Section 3 gives an overview of the image quality metrics used in the experiments. Section 4 gives details of conducted experiments and the results obtained. In section 5 we give conclusions with directions for future work.

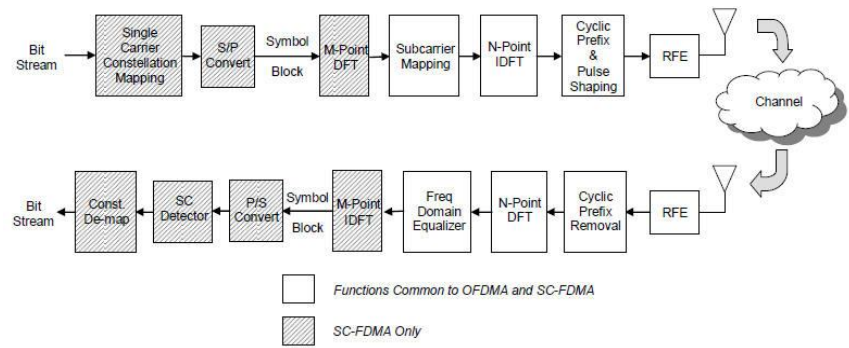

Fig. 1. **Overall block diagram of physical layer in LTE-A**

## 2 LTE PHY details

In this section, we detail about the various parameters in LTE-A, their meaning and significance and also explain the choices we used for implementation for uplink transmission. OFDM is a well-known modulation technique, but is rather novel in cellular applications. The capabilities of the eNodeB and UE are obviously quite different. Not surprisingly, the LTE PHY DL and UL are quite different. The DL and UL are treated separately within the specification documents. Therefore, the DL and UL are described separately in the following sections

Generic Frame Structure one element shared by the LTE DL and UL is the generic frame structure. As mentioned previously, the LTE specifications define both FDD and TDD modes of operation. This paper deals exclusively with describing FDD specifications. The generic frame structure applies to both the DL and UL for FDD operation. The LTE PHY specification is designed to accommodate bandwidths from 1.25 MHz to 20 MHz. OFDM was selected as the basic modulation scheme because of its robustness in the presence of severe multipath fading. Downlink multiplexing is accomplished via OFDMA. The DL supports physical channels, which convey information from higher layers in the LTE stack, and physical signals which are for the exclusive use of the PHY layer. Physical channels map to transport channels, which are service access points (SAPs) for the L2/L3 layers. Depending on the assigned task, physical channels and signals use different modulation and coding parameters.

## 2.1 Turbo Encoder/Decoder

LTE uses a rate-1/3 Parallel Concatenated Convolutional Code (PCCC) consisting of two identical 8-state rate-½ convolutional encoders connected parallel using an internal interleaver. Viterbi decoding of turbo codes is complex due to the large number of states involved in a concatenated trellis. So an iterative MAP (Maximum A Posteriori) detector based algorithm [6] is used as a practical alternative decoding scheme.

## 2.2 Incremental Redundancy HARQ transmission

In the DL, LTE uses asynchronous and adaptive HARQ mechanism. The schedule of the HARQ transmissions is not pre-declared to the UE. This gives the eNodeB flexibility in scheduling according to priorities and available resource. LTE uses Incremental Redundancy (IR) HARQ as opposed to chase combining. LTE supports up to four redundancy versions for IR HARQ (re)transmissions denoted by rvidx = 0,1,2 and 3. In each version, a part of rate-1/3 turbo-encoded data is transmitted dependent on rvidx value.

## 2.3 Rate matching

The rate matching converts the rate-1/3 output from the turbo encoder into the target coding rate. This is done by a block consisting of a three sub-block interleavers, a circular buffer, and a bit-selection block [21]. The number of bits selected depends on the target coding rate. The start point (or offset) of the selected bits is determined by the HARQ redundancy version rvidx.
OFDMA LTE uses OFDMA for DL access. The available frequency is divided into sub-carriers of 15 kHz bandwidth. LTE specific OFDMA parameters are listed are listed in Table 1 [22].

## 2.4 MIMO (Multiple Input Multiple Output)

One of the main features of LTE is the use of multiple antennas or MIMO technology to enhance the throughput in an unreliable wireless channel. A Nt x Nr MIMO system consists of Nt transmitter antennas and Nr receiver antennas. LTE supports MIMO configurations of 4 x 2, 2 x 2, 2 x 1 in the DL. Transmission is done in blocks. Multiple antennas can be used either to obtain more reliable transmissions using Transmit Diversity (TD) or to obtain higher transmission rates through Spatial Multiplexing (SM). In this model open loop MIMO with TD and SM modes associated with 2 x 2 antennas are evaluated in the simulation. In LTE TD is obtained by use of Space Frequency Block Codes (SFBC) as opposed to Space Time Block Codes (STBC).

## 2.5 Channel Modeling

A tapped delay line model is used to model multipath frequency selective channel between transmit antenna and receive antenna [7].

## 3 Quality Metrics Used

The easiest way to measure the quality of video delivered to user is by comparing it with the original and then measuring the quantizations. PSNR or Peak Signal to Noise Ratio is a popular metric for quality assessment. SSIM or Structural Similarity Metric [23-24] is a more accurate measurement of video quality as perceived by the users. But both of them compulsorily need the original video in addition to distorted video for measurements. Such schemes are called as FR or Full-reference schemes. However, in case of mobile phones, when watching a video, users don't typically have the original video. Therefore, FR methods can't be used for such purposes.

Human observers can easily assess the quality of a distorted image without examining the original image as a reference. By contrast, designing objective No-Reference (NR) quality measurement algorithms is a very difficult task. This is mainly due to our limited understanding of the way human visual system behaves. Effective NR quality assessment is feasible only when the prior knowledge about the image distortion types is available.

**Peak Signal to Noise Ratio (PSNR)** metric computes the peak signal-to-noise ratio, in decibels, between two images. This ratio is often used as a quality measurement between the original and a compressed image. The higher the PSNR, the better the quality of the reconstructed image. The Mean Square Error (MSE) represents the cumulative squared error between the compressed and the original image, whereas PSNR represents a measure of the peak error. The lower the value of MSE, the lower the error. To compute the PSNR, the block first calculates the mean-squared error using the following equation:

$$\text{MSE} = \sum_{\{\text{all i}\}} \sum_{\{\text{all j}\}} \bigl(\text{Orig}(i,j) - \text{Rcvd}(i,j)\bigr)^2$$

Then, PSNR is computed using the following equation

$$\text{PSNR} = 10 \log_{10}\left(\frac{255^2}{\text{MSE}}\right)$$

The **structural similarity (SSIM)** index is a method for measuring the similarity between two images. SSIM is designed to improve on traditional methods like peak signal-to-noise ratio (PSNR) and mean squared error (MSE), by considering image degradation as perceived change in structural information. Structural information is the idea that the pixels have strong inter-dependencies especially when they are spatially close. These dependencies carry important information about the structure of the objects in the visual scene. SSIMfirst measure the luminance which is the intensity of the pixel values of the frames. Then, the luminance is removed from the frames and the contrast which is the standard deviation between two frames is obtained. Then the contrast measure is also removed to measure the structural similarity between two frames. The SSIM metric is calculated on various windows of an image. The measure between two windows x and y of common size N×N is:

$$\text{SSIM}(x, y) = \frac{(2\mu_x\mu_y + c_1)(2\sigma_{xy} + c_2)}{(\mu_x^2 + \mu_y^2 + c_1)(\sigma_x^2 + \sigma_y^2 + c_2)}$$

The score varies from -1 to 1, 1 indicating perfect match while 0 indicates no match.

Apart from these, there are other no-reference metrics which don't need the original source video to predict quality.

**Blocking** effect is the most annoying artifact in block transform-coded image/video. Objective measurement of blocking artifact plays an important role in the development, optimization, and assessment of image/video coding systems. It is also very useful for the design and evaluation of the post-processing algorithms at the decoding side.

Thus, we use the blocking metric to evaluate the network degradation caused by LTE network in transmission. In real-time video streaming, done over UDP channels, many parts of or complete video frames are lost leading to distortions or blocking in received video. Also note that blocking metric can be implemented in a no-reference manner so that, we don't need the original copy of the video. Rather, we only need to apply it on a frame to frame basis on the received video. So, we decide to use blocking metrics.

There are many implementation of blocking metrics in research literature [25-28]. [25-26] both use a weighted mean-squared difference along block boundaries as the blockiness measure, where the weights are obtained according to human visual masking effects. However, they cannot distinguish how much of the gray level difference between block boundaries is due to real blocking discontinuity or the oscillation of the original signal itself. [28] proposes a new blind blocking distortion metric by employing the 2nd- and the 3rd-order statistical features of the image. The metric can also be modified to comply with human visual perception by merging the masking effects. This implementation of blocking metrics [28] is most cited in literature and we implemented it in Matlab for our evaluation purposes.

**Blurring** metrics is used to measure the distortions in pixels because of smoothing in the video which can be due to codec losses or packet losses in network [29]. Recently, a completely blind, **Natural Image Quality Evaluator (NIQE)** was proposed [16]. It is based on the construction of a 'quality aware' collection of statistical features based on a simple and successful space domain natural scene statistic (NSS) model. These features are derived from a corpus of natural, undistorted images and fitted into a multivariate Gaussian model.

A natural scene statistic based distortion-generic blind/no-reference image quality assessment model that operates in the spatial domain was proposed in [30,31]. The new model, **dubbed Blind/Referenceless Image Spatial Quality Evaluator (BRISQUE)** does not compute distortion specific features such as ringing, blur or blocking, but instead uses scene statistics of locally normalized luminance coeffcients to quantify possible losses of 'naturalness' in the image due to the presence of distortions, thereby leading to a holistic measure of quality. The underlying features used derive from the empirical distribution of locally normalized luminances and products of locally normalized luminances under a spatial natural scene statistic model. Since these metrics, represent the state of the art, we compare their performance in this work.

## 4  EVALUATION

### 4.1  Experimental Setup

To test the comparative performance of metrics in a wireless transmission setting, we use LTE Physical layer simulator. We bypasss the higher layers such as transport protocol, IP layer, and Network layer in this setup. Rather, we directly connect application layer data to the physical layer, which may seem synthetic but it is justified as our focus is to quantize the impact of different metrics in a practical wireless network setting. The Simulink model proposed earlier by earlier researchers [7,8] was used for the simulations. We report experiments over three video samples – Akiyo video, Rhino video and harbor video to maintain diversity in the sample set. Akiyo sequence is a news-reader reading some content and has mainly lips motion only. Rhino video has motion in parts because of moving rhino in the trace and represents a more natural scene. The harbor video is another natural video but it has motion all around the video because of air movements leading to flag wavering and water moving. These videos are all sampled to size of 320x240 resolution. A screen shot is shown in Figure 2. We also used screenshots of these videos/ standard test images to study the variation in metrics performance with compression ratios. The representative results are shown in this section to demonstrate the trend that we observed.

### 4.2  Experiments

First of all, we vary the channel SNR (Signal to Noise Ratio) from bad to good and stream the videos. We expect that there must be a monotonous variation in quality score of these metrics as video quality is expected to increase with improving channel conditions. As we mentioned earlier, PSNR metrics is ideal for measuring single distortion in the video, although it has been found by researchers to be inaccurate in comparing the perceptual effect of distortions caused by two different things. For example, PSNR can't be used as an effective measure where one image is distorted using Gaussian noise and other distortion is caused by compression-leading–to–distortion. In our scenarios, there is a linear variation in only one parameter, therefore we take widely used PSNR as a reference. We vary the channel SNR from 10 to 20 dB. The modulation was kept constant at 32 QAM while HARQ was set to maximum value. We observe that PSNR values show linear change from left to right in all the videos (Figure 3). It is interesting that widely used SSIM metrics is not useful in this scenario and it shows concave properties (not monotonously increasing as expected).

An ideal metric for real-time assessment of mobile video quality must be able to work in a no-reference manner. Therefore, we next study no-reference metrics and their performance.  In  this  scenario,  all  the  four  metrics  chosen by us, show a monotonous performance and high correlation with PSNR values (see Table 1). However, we can observe that the correlation is highest in case of blocking measure (blocking is measured in dBs) and higher than 99%. Although blurring metrics also show high correlation score, it shows low sensitivity. Although the blurring values vary only between 3.75-4.25 in this case, we observed overall variations in the range of 0-10 for multiple images and multiple distortions. Thus, we can say that the variations or sensitivity of blurring metrics to networking losses is not high.

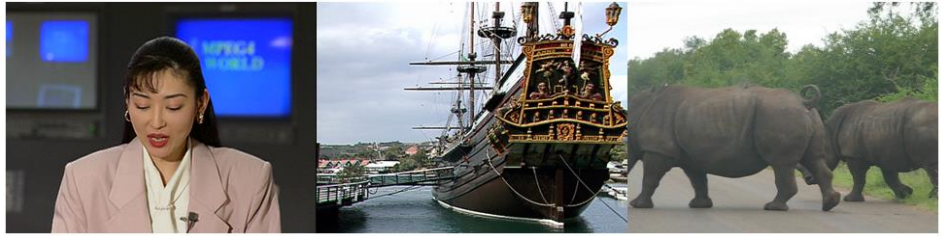

**Fig. 2.** Screenshots of sample video files (a) Akiyo, (b) Harbor and (c) Rhinos

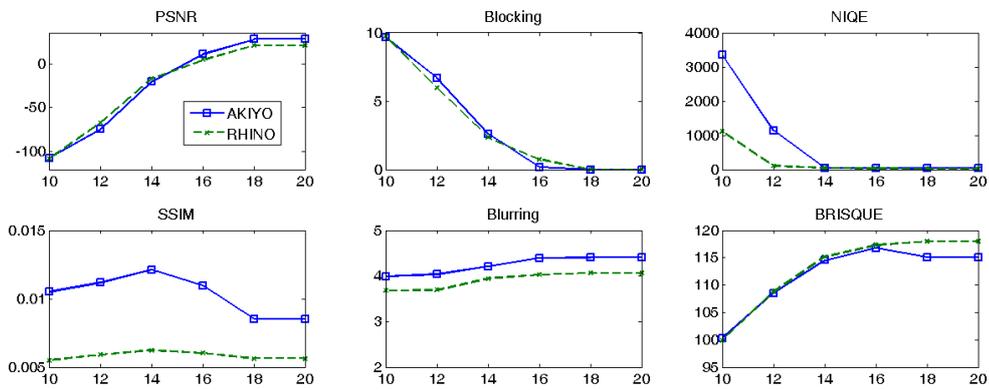

**Fig. 3.** Quality evaluation of different videos streamed over a LTE PHY at different channel conditions (and hence different packet losses). With increased channel SNR (y-axis in each figure), we expect increase in quality of received video quality. This trend is effectively captured by PSNR and all no-reference metrics. Blocking metrics shows highest correlation and sensitivity.

**Table 1.** Correlation of video quality metrics with PSNR values when channel SNR is varied while keeping modulation and code-rate constant. Blocking metrics and BRISQUE metrics give highest correlation.

| Video | SSIM | **Blocking** | *Blurring* | NIQE | **BRISQUE** |
|---|---|---|---|---|---|
| Akiyo | -0.51 | **-0.9936** | *0.9859* | -0.8933 | **0.9273** |
| Rhino | 0.2 | **-0.9969** | *0.9709* | -0.8208 | **0.9821** |

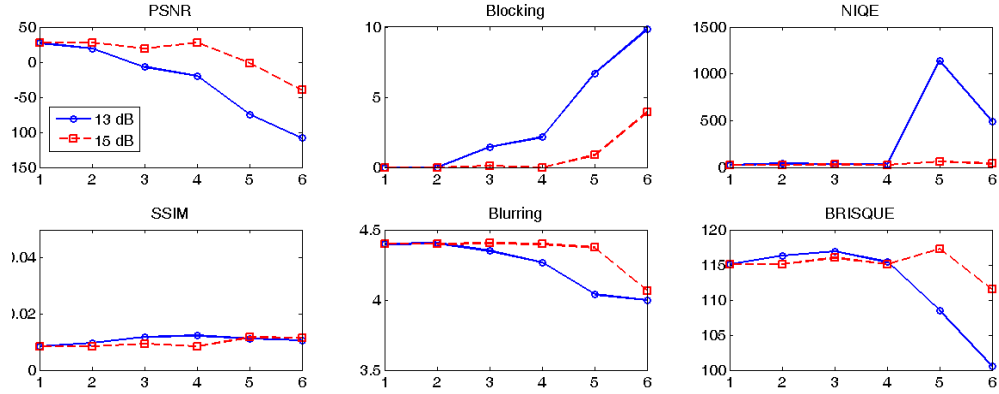

**Fig. 4.** Quality evaluation of different videos streamed over a LTE PHY at same channel conditions (and same different packet losses) but different modulation rates. With increased number of bits per transmission (y-axis in each figure), we expect decrease in quality of received video quality. This trend is effectively captured by PSNR and Blocking and Blurring metrics.

**Table 2. .** Correlation of video quality metrics with PSNR values when modulation rate is varied while keeping channel conditions constant. Blocking and blurring metrics give highest correlation.

| Channel SNR | SSIM | **Blocking** | **Blurring** | NIQE | **BRISQUE** |
|---|---|---|---|---|---|
| 10 db | 0.6985 | **-0.9976** | **0.943** | -0.5335 | 0.8461 |
| 13 db | -0.4139 | **-0.9916** | **0.9844** | -0.7341 | **0.9273** |
| 15 db | -0.8709 | **-0.976** | **0.935** | -0.618 | **0.9821** |

Next, we keep the channel and video to be same and vary the modulation rates. This is plotted in Figure 4. Again, we can see that blocking, blurring and BRISQUE metrics perform well and the scores are highly correlated to the PSNR values. However, Blurring, is not much sensitive and we prefer Blocking and BRISQUE metrics for no-reference quality assessment. Table 2 gives correlation between various metrics with PSNR scores for different channel conditions averaged over videos in our data sets streamed over the LTE network simulator.

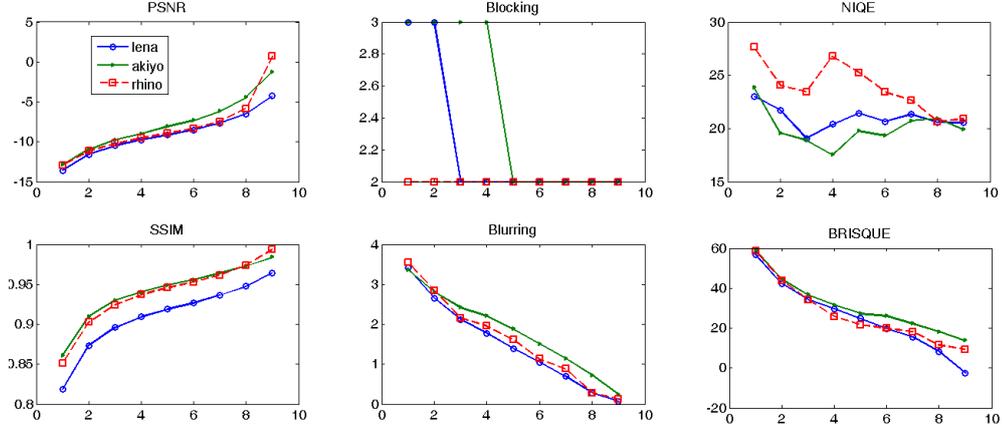

**Fig. 5.** Quality evaluation of different frames compressed using JPEG codec with different compression rates. With increased quality score (y-axis in each figure), we expect increase in quality of received video quality. This trend is effectively captured by all four metrics except blocking and NIQE.

**Table 3.** Correlation of video quality metrics with PSNR values when compression ratio is changed. BRISQUE and Blurring metrics give high correlation.

| Image | SSIM | Blocking | Blurring | NIQE | BRISQUE |
|---|---|---|---|---|---|
| Akiyo | **0.9149** | -0.7808 | **-0.9875** | -0.2043 | **-0.9367** |
| Lena | **0.9594** | -0.7239 | **-0.9768** | -0.4517 | **-0.9959** |
| Rhino | **0.8576** | 0 | **-0.8705** | -0.7646 | -0.8112 |

Another type of distortion is observed when the source coder changes its compression ratio and hence fidelity of transferred video to cope up with network losses. To emulate this effect, we took several video screenshots and images and compressed them with JPEG format. We took the quality level to vary from 10 to 90% in steps of 10%. The results are reported in Figure 5 and Table 3. We found that PSNR, SSIM, Blurring and BRISQUE metrics give good performance to measure this distortion. Particularly, blurring metrics show good performance in this scenario. Blocking metrics doesn't perform that well and gives very poor correlation.

Thus, we can infer that a combination of blocking and blurring metrics can efficiently measure the network degradations in video quality. Another approach is to use the BRISQUE metric alone which is effective in all scenarios.

## 5    Discussions & Conclusions

In this paper, we tried to study the state-of-the-art video quality assessment no-reference metrics for evaluating the quality of mobile videos transmitted over LTE network. We observe BRISQUE metric, proposed recently, to be robust and able to capture all the choice of network degradations in video quality. Blocking metric is able to quantify the degradation caused by packet losses efficiently but not the loss of quality due to source codec. The Blurring metric is able to quantify the losses in quality due to source codec but is not much sensitive to the losses by packet losses.

In our future work, we will try to implement these metrics in a smart phone and test the quality of video streamed by video telephony applications like Skype or Facetime and live video streaming on mobile phones.